# Vcash: A Novel Reputation Framework for Identifying Denial of Traffic Service in Internet of Connected Vehicles

Zhihong Tian, Xiangsong Gao, Shen Su*, and Jing Qiu

*Abstract*—Trust management of Internet of connected vehicles has been a hot topic during the recent years with the rapid development of UGV technologies. However, existing resolutions based on trustworthiness verification among vehicles make the traffic event transmission quite inefficient. In this paper, we assume that the deployed RSUs can provide efficient communication between any pair of RSU and vehicle, and propose Vcash, a reputation framework for identifying denial of traffic service, to resolve the trustworthiness problem in the application level of the Internet of connected vehicles.

In our reputation framework, every vehicle communicates with the RSU directly for traffic event verification, and spread verified traffic event notification. We borrow the idea of market trading, and set up trading rules to restrict the malicious vehicle's spread of false message, and to encourage vehicles to contribute to the traffic event monitoring and verification. To evaluate the effectiveness of our reputation framework, we conduct simulation experiment. Our experiment results indicate that our proposal manages to avoid bogus event spread, and a vehicle in our framework has to contribute to the traffic event detection to normally employ the traffic service.

## I. Introduction

With the rapid development of smart city techniques, a growing quantity of mobile equipment gets involved in people's daily life, bringing great convenience together with huge risk. Vehicles with onboard units are typical examples among such mobile equipment. During recent years, the infrastructure of the vehicular network is developing rapidly, leading to growing interest on the Internet of connected vehicles. However, it is still very hard to identify the malicious information sent from neighboring vehicles[15]. When a vehicle broadcasts a false message in the Vehicular networks, the vehicles in the same network may take the wrong actions, leading to a bad traffic accident. Herein, to construct a robust and secure trust management framework [15-27] for the Internet of connected vehicles turns out to be a critical problem for the actual deployment of UGV.

The reputation-based mechanism is a commonly used technique for the defense of attacks on traffic service. Based on the cooperation of vehicles, researchers evaluate the trustworthiness of traffic warning messages according to the trustworthiness between vehicles (entity-centric methods[15-18]), or according to the message content itself (data-centric methods[22-24]). In spite of the rationality of such two kinds of methods, existing resolutions always involve V2V (vehicle-to-vehicle) network whose trustworthiness is not guaranteed. Thus message broadcast upon such network always needs an additional method to ensure that no bogus events are spread. However, due to the high mobility of the V2V networks, the schemes to ensure V2V network's trustworthiness are costly [22-24].

The existing resolutions assume that the quantity of RSUs (Road Side Unit) is limited [22-30], and V2V network communication is very important to ensure the network connectivity. As a result, the communication between the RSU and the vehicle could be badly delayed, and turns out to be the bottleneck if it involves in the process of verifying the trustworthiness of a traffic event message. However, with the growing deployment of vehicular network infrastructure and the huge demand on the Internet of connected vehicles, the quantity of RSUs is turning out to be sufficient for a quick response to the vehicle request, especially in the downtown area of cities. Thus, we believe that the communication between RSUs and vehicles could be efficient enough for traffic events verification in the near future. Therein, we propose to verify the traffic events based on RSUs, i.e. to ensure all traffic events message sent by RSUs are verified to be trustworthy, and conduct traffic event verification on the RSUs.

Considering the advantages of both entity-centric methods and data-centric methods, and few existing works propose to evaluate the reputation of both entity and data, in this paper, we propose a reputation framework which manages the reputation of both vehicles and the generated traffic events. Our reputation framework is based on an incentive mechanism. All vehicles in our framework are encouraged to contribute to the traffic event detection as much as possible. For each vehicle, the detected traffic events have to be correct, or the vehicle would be punished. To that end, each vehicle is initialized with a certain amount of capital (noted as vehicle cash), and a vehicle has to invest on each traffic event it spreads to make it accepted by the RSUs. The RSUs sell the traffic event for the generator if the traffic event could be verified. Thus the invested traffic events

Zhihong Tian, Shen Su, and Jing Qiu are with the Cyberspace Institute of Advanced Technology, GuangZhou University, GuangZhou, China

Xiangsong Gao is with China Academy of Engineer Physics, Mianyang, Sichuan, China.

Corresponding author: Shen Su, Email: johnsuhit@gmail.com

could be profitable if it is a de facto traffic event. However, for malicious vehicles which send bogus events, they would quickly run out of their vehicle cash, and could send no more bogus events since none of their generated traffic events are profitable. For vehicles which send no messages, they would also run out of vehicle cash, and could employ the traffic service no more since every traffic event warning costs.

Our proposed reputation framework could be taken as a combination of the entity-centric methods and data-centric methods. Every vehicle's Vcash indicates its capability to spread invested traffic events, thus could reveal the reputation of the vehicle. The investment of each traffic event is used to decide whether the RSUs should accept the traffic events, thus could reveal the reputation of the data, and determine the management conducted to the data.

Our contribution in this paper includes three folds:

First, we propose vehicle cash, a reputation framework for the identifying denial of traffic service, which applies for the trend of growing deployment of RSUs, and growing interest in the Internet of connected vehicles.

Second, we discuss the multiple threaten modes for the denial of traffic service attack, including the selfish attack mode. We also propose a corresponding resolution to restrict the spread of false message, and to encourage contribution of traffic condition monitoring and verification.

Third, we conduct simulation experiments to prove the effectiveness of our reputation framework. Our experiments indicate that Vcash could efficiently reduce the bogus event spread, and encourage all vehicles involving in traffic event monitoring and verification.

In the rest of this paper, we will first introduce the related works in Section2. Then will give the problem statement in Section 3, and we present our reputation framework in Section 4. We evaluate the effectiveness of our reputation framework in Section 5, and conclude our proposal with future work in Section 6.

## II. Related Works

The security issue is considered as the very fundamental of the Internet of connected vehicles, and numerous outstanding works have been devoted to this field. The foundational security aspects mainly concern the behaviors that would threaten the functionality of Internet of connected vehicles, researchers have proposed several schemes and frameworks to cope with the corresponding problems. Refs. [1] proposed the mobile protocols in order to promote the effectiveness of communication between the vehicles. [2] designed the layer protocols and control mechanism to improve the tracking accuracy as well as decrease the congestion of vehicular networks. These researches also consider problems such as the information aggregation [3], DDOS attack [4], the trade-off between security and performance [5], et al.

The other works mainly contribute to the security of the application level, and these works have been well surveyed in [6-14]. In general, the dishonest and the refuse-to-participate behaviors are considered as the most severe threat to the vehicular networks, and the reputation mechanism is widely adopted to conquer these threats. Two types of arbitration methods are included in the state of the art contributions.

The entity-based researches such as [15-18] consider the vehicle as the object of the reputation, and they will decide the behavior of vehicular networks by judging the reputation of each vehicle entity.

Refs. [19] proposed an evaluation method for the data generating process. Every single vehicle is set up with a processing flow of behaving and validation. This design of reputation is also involved in its future researches like [20] and [21]. However, most of these researches are considered as low real-time performance [7].

In other researches, the event-based solutions such as [22-24] tend to assign the reputation value to the runtime event. Refs. [25] proposed an event-centric trust establishment framework and applied it to the traffic safety application. The novel concept is to evaluate the trustiness of sensed data or received messages rather than the trust of the individual vehicle. Beside [25], other researches which tried to optimize the method generally attempted to arbitrate by the time [23], sensor capability [24] and so on. However, the black hole attack [26] cannot be avoided in such kind of frameworks. [27] proposed an incentive-based method to conduct credit management in the vehicular networks, which rewards positive activity and punish negative activity. Their method also suffers from the vehicle-to-vehicle trust management in spite of their incentive-based credit management scheme. As a wider range, wireless issues has also drawn much more research interest [28-35] during the last two decades.

## III. Problem Statement

Generally, our idea in this paper is to build a reputation based trust management framework for the Internet of connected vehicles. Since the end-to-end communication security is not our topic in this paper, and the message confidentiality and integrity could be ensured by encryption and digital signature techniques, in this paper, we assume that all vehicles take necessary approaches to ensure the end-to-end message communication security. As a result, we discuss our foundation of reputation framework in the application layer, more specifically, as the foundation of traffic service.

The objective of our reputation framework is to encourage all vehicles in the network of connected vehicles to provide qualified sensed traffic events. Herein, we focus on identification of the malicious vehicles which keeps sending bogus events or send no message at all.

In the rest of this section, we will present the basic assumption and statement of our problem.

### A. Network Model

In our vehicular network which provides traffic service, the communicating entities are vehicles (more specifically, the OBU, onboard units) and base stations on the roadside (i.e., RSU). Considering the vehicles and RSU could involve in different applications, we assume that all vehicles and RSUs could communicate with each other. An OBU has limited communication range, thus could only communicate with a limited number of other OBUs and RSUs. All OBUs and RSUs take appropriate approaches to ensure the information integrity and confidentiality.

Vehicles run in bi-direction on the road with different speed, and every vehicle is equipped with appropriate sensors to capture the traffic events (such as traffic jam, traffic accident, road condition, etc.). In our network model, traffic events randomly happened in the roadmap, and last for a certain period of time. We assume that the vehicles appropriately cover the traffic events in our roadmap, i.e. all traffic events would be efficiently verified in our network.

The sensed traffic events information (including time, location, traffic event type or description, etc.) are collected by the onboard units, and sent to other OBUs and RSUs. All vehicles travel randomly. With the appropriate warning of the traffic service, a vehicle would be able to avoid running into any traffic events by surprise.

### B. Threaten Modes

Since we assume secure end-to-end communication, to conduct denial of traffic service, a vehicle could: 1) send a bogus event including fake traffic events; 2) or send no messages at all.

- Bogus event mode

If a vehicle keeps reporting fake traffic event messages, we term that vehicle is in a bogus event mode. A fake traffic event message refers to a message indicating traffic events which never happen, or at wrong time/location. If a vehicle keeps receiving this kind of message, it could be a disturbance to the driver. And such fake warning could bring down the trust of the driver to the traffic service. More importantly, when a genuine traffic event happened and sent to the passing-by vehicles, such waring may draw very little attention because of the low trust of the traffic service, leading to potential driving risk.

The bogus event mode applies to the malicious vehicles which intentionally generate fake messages and broadcast them in the network. Moreover, the bogus event mode also applies to the vehicle with unqualified sensors or software. Due to the defects of hardware and software, a vehicle could generate inaccurate information and tell other vehicles wrong information.

- Selfish mode

If a vehicle reports no traffic event message at all, we term that vehicle is in a selfish mode. The bogus event mode indicts sending fake messages, i.e., false positive detection in the traffic service. To the opposite, false negative detection could also decrease the drivers' trust in the traffic service, and surprise the driver when they encounter the traffic events.

Since assurance of qualified traffic environment sensing could increase the vehicle maintenance cost, a vehicle has the motivation to shut down all traffic event sensing and use the traffic service. However, if all vehicles work in such a manner, the quality of traffic service could still be terrible. Herein, we need to discuss how to defend from this attack mode, and encourage all vehicles to send qualified sensed traffic events.

### C. Traffic Events Detection and Verification

In this paper, we assume all traffic events are sensed by vehicles without the help of RSU. We make this assumption because sensors can only accurately cover a limited ranage; thus an RSU can merely cover the traffic events within a limited range. However, vehicles can cover all over the roadmap since their mobility. As a result, the help of RSU is very limited, and we do not want to make the problem more complex with the involvement of traffic events captured by RSUs.

When a traffic event happens, it would be detected by the first vehicle capturing its corresponding information and broadcast it all over the network (if it is not an attacking vehicle). To verify the existence of the reported traffic events, we rely on the cross-validation of the reported events, i.e., if a traffic event is reported multiple times, there would be a high possibility that it is a de facto traffic event.

### D. Problem Formulation

Given the above assumptions and descriptions, our problem could be formulated as an optimizing problem defined on a 5-tuple (*V*, *RSU*, *E*, *C*, *f*).

- *V:* The set of vehicles in the network. A vehicle $v \in V$ is represented by a 5-tuple (*id, location, dir, vel, event(id)*). Here *id* refers to the identification of the vehicle; *location* refers to the vehicle's location; *dir* refers to the vehicle's direction; *vel* refers to the vehicle's velocity. $event(id) = (e_v^{(1)}, e_v^{(2)}, ..., e_v^{(n)}, ...)$ refers to the reported traffic event messages in time sequence. A vehicle could work in a normal mode, which generates qualified traffic event reports; a bogus mode which keeps generating fake traffic event reports; or a selfish mode which generates no traffic event report.

- *RSU:* The set of roadside units. The information of a roadside unit includes its location and identification.

- *E:* The set of traffic events happening in the problem roadmap. We use a 4–tuple (*start, end, location, attr*) to represent a traffic event e. Here *start* refers to the start time of a traffic event; *end* refers to the end time; *location* refers to where the event happens; and *attr* refers to a set of certain predefined event attributes (e.g., traffic jam, overflow pavement).

- *C:* a successive value range which is a metric of the trust of a vehicles or a traffic event message. A higher value represents more trust.

- *f:* a mapping from $V \cup E$ to *C*. Obviously, *f* assigns a reputation to each vehicle and event.

As a formulation, we aim at optimizing the mapping of *f*, for de facto events, the corresponding vehicles should learn the messages with high credibility; for fake events or outdated events, the corresponding vehicles should learn the messages with low credibility or ignore them. As for vehicles, malicious vehicles or selfish vehicles should not be able to generate high credibility events. And normal vehicle should be able to keep sensing the traffic condition to work for the traffic service.

## IV. REPUTATION FRAMEWORK

Both entity-centric reputation framework and event-centric reputation framework have their intrinsic advantages, and we believe that both the communication entities and traffic event messages should be evaluated by a corresponding reputation for trust management on the Internet of connected vehicles. The event reputation is needed since traffic event is always dynamic. Entity reputation is needed since threatening vehicles always behave similarly because of their original intention, and recognizing such entities could improve the accuracy of traffic event evaluation. Herein, at the end of the trade-off between entity-based methods and event-based methods, we propose VCash (Vehicle Cash) as a reputation framework for trust management on the Internet of connected vehicles.

Our inspiration comes from the market system. In a market system, an entity has to carefully invest in the right product to ensure the normal running of its business. Otherwise, the entity's cash flow would fracture if the invested product cannot be sold. Similarly, we make our reputation framework works as a market, all vehicles involved act as a business entity and initialized with a certain amount of vehicle cash. When a vehicle captures a traffic event, it has to invest a certain amount of vehicle cash on the traffic event. Thus, a vehicle in our framework has to invest in the successful events to ensure its vehicle cash flow could normally run. By setting up trading rules, our market guarantees: 1) merely the de facto traffic events are profitable; 2) traffic events invested with little vehicle cash could hardly be accepted. As a result, all traffic events broadcasted by vehicles working in bogus event mode of selfish mode can hardly be approved.

Our reputation framework utilizes vehicle cash to represent the reputation of vehicles and broadcasted traffic events. The vehicles owing more vehicle cash could broadcast more high reputation traffic events. The traffic events with more vehicle cash indicate the detector is in good marketing condition. In the rest of this section, we will first give an overview of our framework, followed by the framework detail including vehicle service model, zoning market and trading plans.

### A. Framework Overview

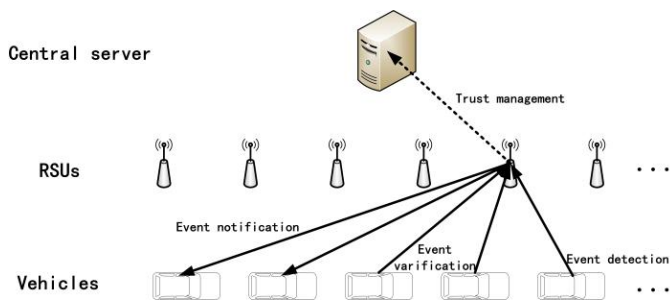

Figure 1 Framework infrastructure

We first introduce the overview of our reputation framework. On the point of infrastructure (shown in Fig 1), our reputation framework is composed of vehicles, RSUs, and central servers. The vehicles sense the traffic condition, capturing traffic events, and send the sensed events to the RSUs. To benefit from the traffic service, vehicles also need to capture the broadcasted traffic events sent by other vehicles. Since our reputation framework works on the base of trading, we need a trustable trading platform to conduct event investment and event selling. In this paper, we assign the job of trustable event trading to the RSUs. I.e., we assume that the RSUs are trustable in this paper. When a vehicle detects a traffic event message, it sends the event message to the RSUs for event verification. When the detected traffic events are verified according to the method discussed in Sec 3.C, the RSUs then broadcast the traffic event message to the other vehicles. Considering the mobility of the vehicles, an RSU may need to deal with the trading of vehicles out of its range limit. Thus we need a central server to take charge of communication of multiple RSUs. The central server collects the result of event trading, and maintains a global view of each vehicle. The central server may encounter a scalability problem. However, such a problem involves the distributed computation framework which is also a non-trivial problem. Since we do not want to discuss another non-trivial problem in this paper, our framework involves a single central server. The following works may introduce distributed computation techniques to our framework to solve the scalability problem.

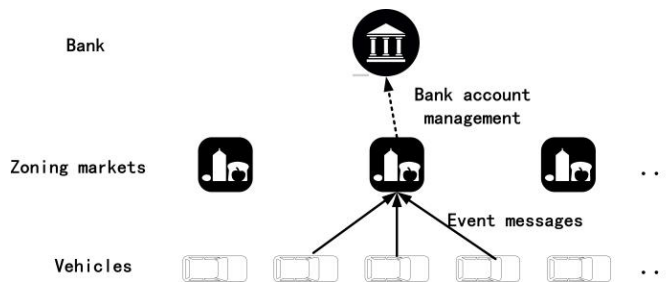

Figure 2 Framework functionality

On the point of functionality, our reputation framework is composed of vehicle clients, zoning markets, and the bank (as shown in Fig 2). Vehicle clients are client applications running on the vehicles with an OBU. When a vehicle client detects a traffic event, the vehicle client pushes the event into the corresponding zoning market for event verification and selling. A zoning market is a server working on one or multiple RSUs, and manages the traffic events within a certain area. Thus a zoning market collects the traffic events happening in its response area, and charges the vehicles traveling in the area for notification of such events. Since a vehicle may travel from area to area, a vehicle may trade events in multiple zoning markets. Thus we need to manage the vehicle cash in a bank account for every vehicle client. In this paper, we manage a vehicle cash bank on the central server for bank account management.

Our framework infrastructure is very similar to the existing works since they all conduct traffic service. Since our proposal is mainly about traffic event trading, in the following we introduce the detail of our reputation framework on the point of functionality.

### B. Vehicle Client

As shown in Figure 3 is the structure of our vehicle client composed with event generator, event listener, event handler, event table, and event sender. The event generator takes the streaming data generated by vehicle sensors as input and produces corresponding traffic event information (event attributes, timestamp, and location). The generated event information is directed to the event handler.

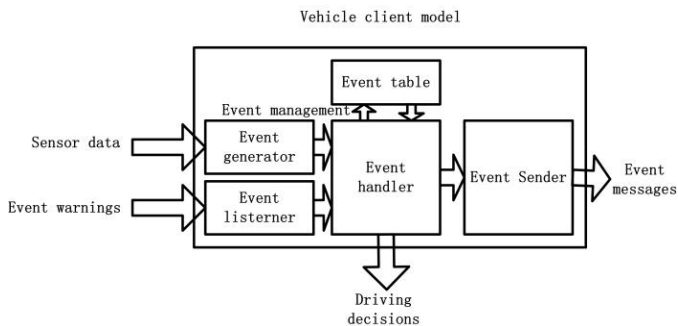

Figure 3 Vehicle Client model

The event listener monitors the event notifications which indicate the occurring traffic events. The event notifications are sent to the vehicle client, when the vehicle runs into the coverage area zone, when a new verified traffic event is broadcasted, or when a traffic event is verified to finalize.

The event handler plays a central role in the vehicle client, since all the generated events and event notifications are directed to the event handler for its driving warning and decision. When the event handler receives a generated event, it integrates the vehicle identification into the event message, looking up the event routing table, and sends the event to the zoning market if the event is unknown. When the event handler receives a traffic event notification, it manages the event table according to the notification. An event notification could be event announcement of event withdraw; thus the management could be an insertion or delete of the event table. The event handler also monitors the occurrence of the existing traffic events in the zone, and if the vehicle passes by the location where an existing traffic event happens, and no traffic events are captured, the event handler will send a message to the zoning market to report the non-existence of the traffic event. When the vehicle runs out of a zone, the event deletes all the events in the event table.

The sender directly communicates with the zoning market deployed on the neighboring RSU through the V2V or V2I network communications.

It is noted that we do not conduct event investment in the vehicle client. Instead, we manage the account of each vehicle direct in the zoning market deployed on the RSUs. We do this because the event trading is deployed on the zoning market. Considering that the vehicle bank account is deployed on the vehicle client, an event trading has to wait for the confirmation of high-speed moving vehicles, and the trading efficiency would be badly delayed.

### C. Zoning Market

A zoning market is responsible for the traffic event management of a certain block area (i.e., a zone), and deployed upon a set of RSUs within the block area. The main job of the zoning market is to: 1) conduct traffic event lifetime monitoring and verification; 2) and conduct event trading by the management of vehicle accounts according to the predefined trading plans.

In Figure 4, we describe the process of communication between the vehicle, the zoning market, and the bank during the period that the vehicle gets through the zone.

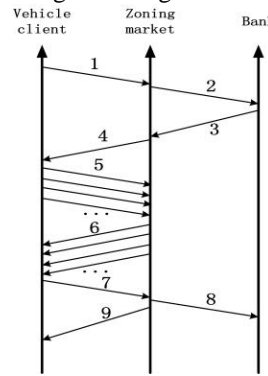

Figure 4 Communication process between zoning market, vehicle, and the bank

When the vehicle firstly gets into the zone, it sends a message including its identity to the zoning market (①), then the zoning market sends a message to the bank to request the bank account information of the vehicle according to the identity of the vehicle (②). The bank sends back the account information if the corresponding account exists, or else creates a new account for the vehicle (③). Then the zoning market sends an acknowledgement message back to the vehicle to start trading events for the vehicle(④). When the vehicle travels in the zone, it monitors the traffic condition and sends corresponding messages to the zoning market (⑤). The zoning market correlates the received event messages, if multiple vehicles send similar event messages to the zoning market (similar event attributes, location, etc.), the zoning market will verify the existence of the traffic event with appropriate charge, and broadcast the traffic events to the other vehicles in the zone (⑥). When the vehicle runs out of the zone, it sends a message to the zoning market (⑦), making the zoning market upload its account information to the bank (⑧). After the vehicle gets out of the zone, since the traffic events generated by that vehicle may still exists in that zone, the zoning market keeps managing the profit of the corresponding events(⑨). It is also noted that ⑧ may happen after ⑨), because the event may keep earning vehicle cash with the event left in the zoning market. An event would stop training if it is verified that the event no longer exists in the zoning market.

### D. Trading Plans

In the above communication process, we set up the following rules to restrict the broadcast of event messages sent by malicious vehicles, and to encourage contribution to the traffic service.

*1) Event investment and profit*: Every event announcement and event verification has to be invested in our system. The investment would get profit if the announced event keeps alive. When the event terminates, all the investment goes to the vehicle which verifies the non-existence of the event. However, if an event verification if proved to be wrong, its investment would be expropriated.

In step ⑤, the zoning market invests a certain amount of vehicle cash from the vehicle account into the event according to a predefined trading plan. If the invested vehicle cash is less than a predefined threshold, the zoning market will ignore the event message, i.e., if a vehicle account has little cash left, the messages sent by that vehicle can hardly be accepted by the zoning market. If multiple vehicles invest on the same event (same event location and event attributes), the zoning market grants each vehicle a share of the event stock according to each vehicle's investment of the event, and distribute the earnings of each event according to each vehicle's stock (according to formula 1).

$$S(v_i) = \frac{I(v_i)}{\sum I(v_i)} \times c \times |ZV| \qquad (1)$$

Here $S(v_i)$ refers to the profit sharing of the vehicle $v_i$; $I(v_i)$ refers to the investment of $v_i$; $\sum I(v_i)$ refers to all the investment on the profiting traffic event; $c$ refers to the fee for a traffic event notification which is a constant number; and $|ZV|$ refers to the set of vehicles paying for the event.

To the end of protecting the broadcast of malicious event messages, we need to ensure that each vehicle cannot send a lot of event messages to the zoning market before its invested events start earning a profit. In this paper, we plan to invest a certain percentage of the rest vehicle cash in the vehicle account (as shown in formula 2).

$$I(n) = V(n-1) \times ratio \qquad (2)$$

Here $I(n)$ refers to the invested vehicle cash of the $n^{th}$ event, and $V(n)$ refers to the rest of vehicle cash after investing on the $n^{th}$ event. Obviously, $V(n) = V(n-1) - I(n)$. And *ratio* refers to the constant ratio we invest in each iteration.

*2) Event charging:* In steps ④ and ⑥, the zoning market charges each vehicle client a certain amount of vehicle cash for each traffic event notification, which actually refers to the parameter $c$ in equation 1. If the vehicle client doesn't have enough money for all traffic events notification, the zoning market ranks all the events in a descending order according to their investment, and sends as many traffic event notifications as possible until there is not enough vehicle cash for more traffic event notification. In addition, the charge rate varies according to the number of events in the system. Simply put, the more events, the lower the rate. It is also noted that, with the restriction on the charging rate, we would be able to punish the selfish vehicles, since selfish vehicles would not be able to pay for the event notification because they has no way to earn money in our system.

*3) Event termination:* If multiple vehicles send event messages to verify that a traffic event is nonexistent, the zoning market would terminate and the event, then take the vehicle cashes invested on the event as reward with the other vehicles which contribute in the event termination (according to formula 3).

$$S(v_i) = \frac{I(c_i)}{\sum I(c_i)} \times \sum I(v_i) \qquad (3)$$

Here $S(v_i)$ refers to the profit sharing of the vehicle $v_i$; $I(v_i)$ refers to the investment of $v_i$; $\sum I(c_i)$ refers to all the investment on verification of the event non-existence, and $\sum I(v_i)$ refers to all the investment of the event.

## V. EVALUATION

In this section, we discuss the effectiveness of our framework based on simulation. We discussed two attacking mode in this paper: selfish mode and bogus event mode. For the selfish mode, since vehicles in our framework has to pay for event notification, a selfish vehicle would be kicked out of the framework if it contribute nothing to the framework. In the rest of this section, we will focus on the protecting performance of our framework with all malicious vehicles working in the bogus event mode.

### A. Simulation setup

To evaluate the effectiveness of our proposed framework, we develop an event-based simulator, which simulates parallel vehicle behaviors. Our simulator takes self-defined roadmap, vehicles, and events, and we use it to simulate the vehicular network scenario happens in a zoning market. Since we assume that there are enough RSUs to ensure the efficient vehicle-RSU communication, we consider no channel status changing problems in this paper.

In our simulation, the road map is a 5000 meter long ring road, which simulate the high way. We randomly generate and place a quantity of vehicles on our ring road. Every vehicle runs at a random speed between 10 and 30 meters per second with a random initialized direction. To simplify our problem analysis, we assume that the road of our experiment includes only one zoning market.

We set up 10 events randomly distributed on the road at the beginning of the experiment. The duration of these events is randomly distributed between 60 and 600 seconds. After everything starts to work, every 60 seconds will have new events happening randomly on the road. Its duration is consistent with those of the initial events. Our simulation lasts for 1000 seconds. We rerun our simulation for 50 times, and calculate the average result as our evaluation result.

We also place a number of threatening vehicles in our simulation. Since the main inaccuracy is caused by the vehicles works in bogus event mode, we focus on the evaluation of bogus event mode vehicles. It is noted that if our traffic event verification scheme needs more than two nodes to verify the existence of traffic events, and the malicious vehicles send a random bogus event, the bogus event can hardly be spread. Therein, we assume that all bogus event vehicles collide, and share one false event map. All such vehicles report a false traffic event every second. We use this as the "event flood" attack mentioned in the previous article. In our simulation, each vehicle is initialized with 100 vehicle cash. Since we want to set up a balanced charge rate in our simulation, the charge rate of each event notification should be more or less equal to the investment of a traffic event. Therein, based on the floating rates mentioned above, we set the initial rate ("$c$" in equation (1)) to one in ten thousand per event per second. As the number of events in the system changes, the specific rate will be the initial rate divided by the number of events. To accelerate the working process of our evaluation framework, we accelerate the bogus events emission speed as 1 bogus event per second, so that the malicious would be quickly recognized.

### B. Traffic service accuracy

Our reputation framework aims at restricting the bogus event spread. Herein, our key metric to evaluate the reputation framework is the amount of bogus events compared to the normal messages. In Figure 5, we show the percentage of bogus events out of all existing events over time in our simulation. As described in the previous subsection, our simulation starts from a randomly initialized status, and we rerun the our simulation for 50 times and show the average result value in Figure 5. Considering that the amount of randomly placed vehicles in our simulation would impact the experiment results, we conduct 3 groups of experiments with various vehicle placement density (10 vehicles per kilometer, 20 vehicles per kilometer, and 30 vehicles per kilometer). Considering our framework requires multiple messages to verify a bogus event, we also conduct our experiments with various "m" parameters, which refers to the quantity of traffic event messages needed for verification of traffic event termination.

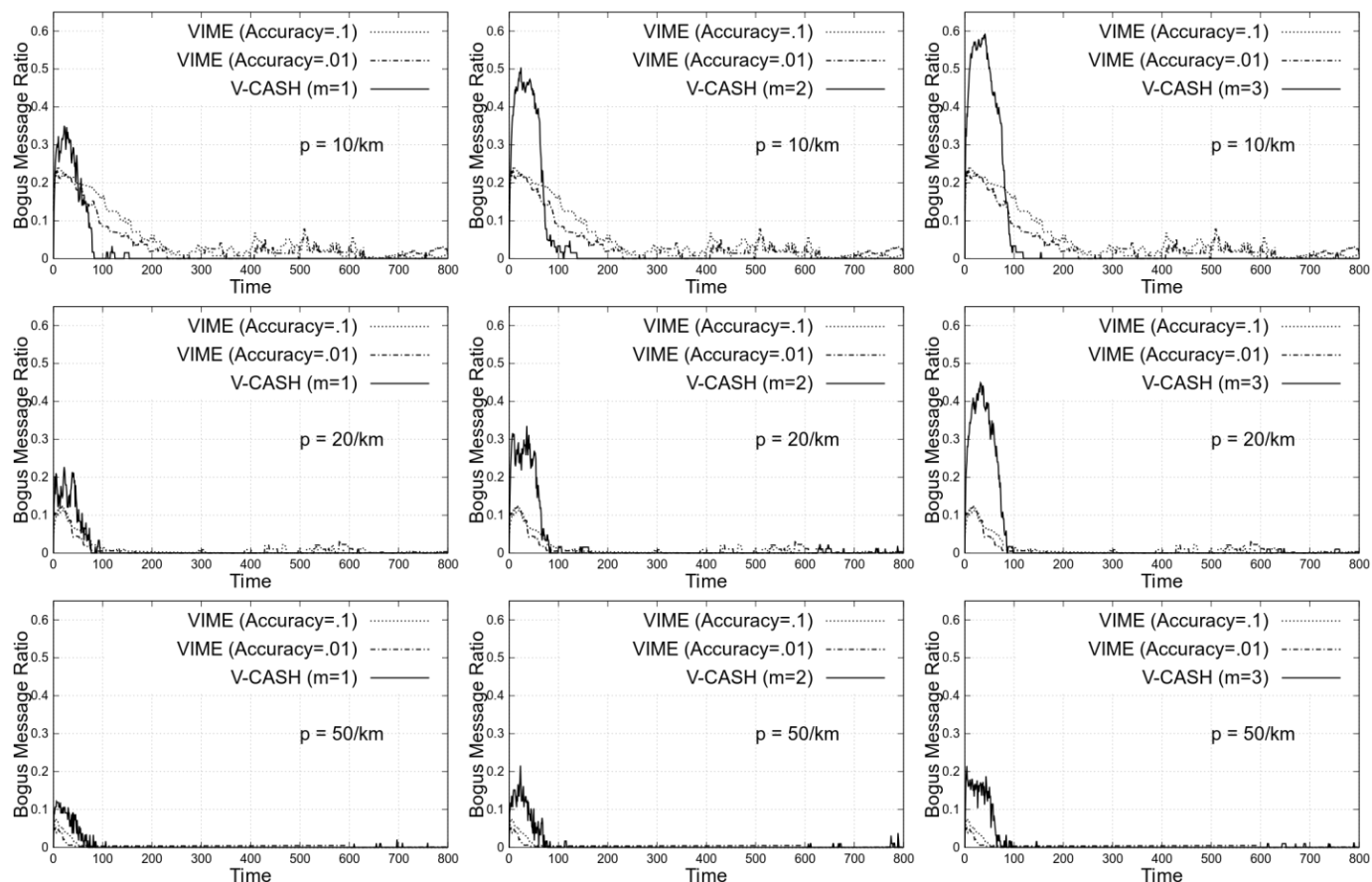

**Figure 5.** Bogus event ratio over time (seconds). We compare our method with VIME used in [27]. Since the performance of VIME relies on the accuracy their entity based reputation, we set the accuracy as 99%( Accuracy=.1) and 99.9%( Accuracy=.01) individually. $p$ refers to the density of vehicles on the road map which is 10 vehicles per kilometer for the 3 figures in the first line, 20 vehicles per kilometer for the second line, and 30 vehicles per kilometer for the third line.

As a comparison, we also conduct the reputation framework proposed VIME [27], which is similar to the proposal of our work, but managing individual entity-based reputation. Considering that VIME conduct individual reputation evaluation based on gathered trust information which we actually do not simulate. We simply give a chance (99% and 99.9% individually for curves "VIME(Accuracy=.1)" and "VIME(Accuracy=.01)" in Figure 5) to VIME to make the accurate entity-based reputation evaluation, which is fairly accurate compared to existing works.

As observed in Figure 5, both methods' effectiveness detects an explosion of bogus events, then the involved malicious vehicles got recognized with no more detected bogus event. This is not surprising because all malicious vehicles have enough credit to invest on the bogus events at the beginning of our simulation. After the explosion, most of spread bogus events are recognized with no pay back. And the malicious vehicles turn out to be silent. For our method, the explosion lasts for about 100 seconds, with barely no tail (no bogus event appears after 100 seconds). However, VIME (reputation accurate ratio =99%) has quite a long tail. Even at the end of our simulation, VIME's bogues message ratio is still non-trivial.

Considering one may argue that the long tail is to blame the inaccurate bogus event recognition, we improve the reputation accurate ratio to 99.9%, which makes the tail turns to be thinner. However, the tail still exists during our simulation. Compared to VIME, our method suffers from the explosion, but manages to converge much faster. Because of the assumption that most vehicles manages to recognize the bogus event once upon receiving the event, our method suffers from a bigger bogus event explosion. However, since VIME conducts peer-to-peer reputation, the spread of malicious vehicle recognition turns to be much slower than our method.

As we utilize multiple events to cross validate a bogus event, the more events needed for cross validation, the bigger explosion we would encounter at the beginning of our simulation. At the same time, vehicle placement density also impacts our simulation result, the more vehicles placed on the roadmap, the smaller scale of explosion our simulation suffers, because that more vehicles would help the recognition of bogus events.

To sum up, with appropriate and careful verification of traffic event occurrence and termination, we believe that we are able to restrict the spread of bogus event in a rational range limit, which converges much faster than existing works.

*C. Vehicle cash*

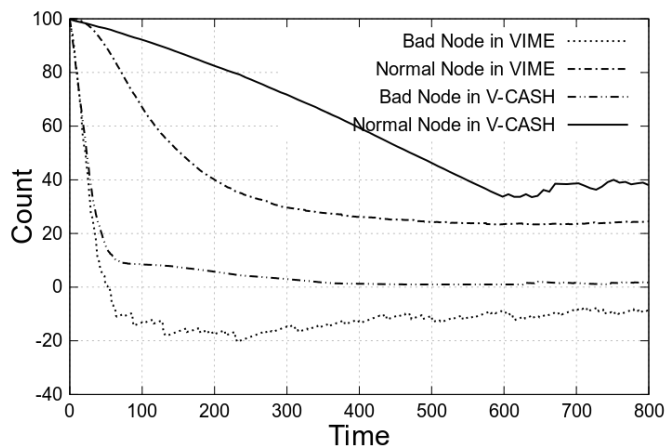

**Figure 6.** Vehicle cash of the bogus event node over time (seconds). The y-axis represents the amount of vehicle cash in each vehicle account. "Bad node" refers to malicious vehicle in VIME and V-CASH.

Another important metric is the system investment on traffic events. How much Vcash invested on the traffic events indicates the confidence of the vehicles declare the event, which could be useful for the traffic service users. The problem is that, each vehicle needs to have enough Vcash to conduct investment. Herein, we study the Vcash changes (reputation credit of each vehicle) overtime in our simulation as shown in Figure 6.

In Figure 6, with no surprise, for both methods, normal vehicles keep at a much more higher level than the malicious nodes, which makes the normal vehicles able to spread traffic messages when detect one, but the attacking vehicle can hardly send messages because their Vcash gets to very little amount after a few rounds of bogus event investment.

The Vcash of normal vehicles for our method shows a trend to gather a arising amount of Vcash, indicating that our method would fairly reward the proper traffic event declaration. However, reputation credit of VIME's normal vehicle keeps at a stable standard, which is much lower than the initial amount.

## VI. CONCLUSION AND FUTURE WORKS

In this paper, we propose Vcash, a reputation framework for identifying false traffic event messages on the Internet of connected vehicles. By means of event trading between multiple vehicles in the zoning market with appropriate trading plans, our framework is able to avoid the bogus event spread, and encourage the vehicles to contribute to the process of traffic event detection. According to our simulation evaluation on highway scenario, we prove that our reputation framework could manage to punish the malicious vehicles which spread a bogus event or contribute no traffic detection. By increasing the quantity of traffic event verification, we can appropriately decrease the amount of false positive rate and false negative rate in our reputation framework. As future research direction for the improvement of our work, we would apply more roadmap and traffic scenarios to our framework including the multiple zoning markets scenario, and investigate the appropriate and dynamic charging rate setup to satisfy the dynamic zoning market trends.

ACKNOWLEDGMENT

This work was supported in part by the Guangdong Province Key Research and Development Plan (Grant No. 2019B010137004), the National Key research and Development Plan (Grant No. 2018YFB0803504 and No. 2018YFB1800701), and the National Natural Science Foundation of China (Grant No. 61902083).

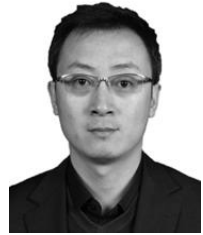

**Zhihong Tian,** Ph.D., professor, PHD supervisor, Dean of cyberspace institute of advanced technology, Guangzhou University. Standing director of CyberSecurity Association of China. Member of China Computer Federation. From 2003 to 2016, he worked at Harbin Institute of Technology. His current research interest is computer network and network security. E-mail: tianzhihong@ gzhu.edu.cn.

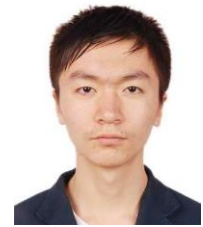

**Xiangsong Gao**, PH. D candidate, China Academy of Engineer Physics. His current research interest is game theory and network security.

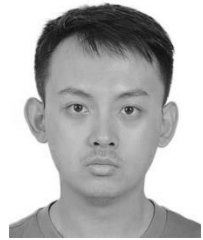

**Shen Su**, born in 1985, Ph.D., assistant professor, Guangzhou Unversity. His current research interests include inter-domain routing and security, Internet of connected vehicles, and wireless sensor networks. E-mail: johnsuhit@gmail.com.

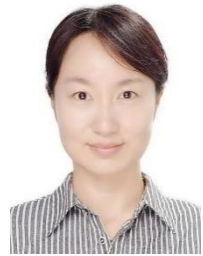

**Jing Qiu,** received the Ph.D. degree in computer applications technology from Beijing Institute of Technology. She was a Visiting Scholar with the University of Southern California, LA, USA, under the supervision of Professor Craig A. Knoblock. Her current research interest is Information Extraction, Network Representation, and Big Data Analysis. E-mail: qiujing@ gzhu.edu.cn.